\documentclass{emulateapj}
\usepackage{amssymb}

\begin{document}
\title{Astronomy, Astrophysics \& Space Physics in Greece}

\author{Vassilis Charmandaris\altaffilmark{1}}

\affil{University of Crete, Department of Physics, P.~O. Box
  2208, GR-71003, Heraklion, Greece}
\email{vassilis@physics.uoc.gr}

\altaffiltext{1}{Chercheur Associ\'e, Observatoire de Paris, F-75014,
  Paris, France}
\slugcomment{Paper to be published in ``Organizations and Strategies in
Astronomy -- Vol. 7, Ed. A. Heck, 2006, Springer, Dordrecht."}

\begin{abstract}

In the present document I review the current organizational structure
of Astronomy, Astrophysics and Space Physics in Greece. I briefly
present the institutions where professional astronomers are pursuing
research, along with some notes of their history, as well as the major
astronomical facilities currently available within Greece. I touch
upon topics related to graduate studies in Greece and present some
statistics on the distribution of Greek astronomers. Even though every
attempt is made to substantiate all issues mentioned, some of the
views presented have inevitably a personal touch and thus should be
treated as such.

\end{abstract}

\section{Introduction}

The framework within which astronomers -- a term that will be used
rather loosely in the rest of the document to indicate individuals
performing research in Astronomy, Astrophysics and Space Physics
(AA\&SP) -- have been functioning in Greece is not too different from
other European countries.

As in most other countries in Europe, the educational and research
activity in Astronomy, Astrophysics and Space Physics in Greece has
been fostered within public Universities and Research Institutes. Even
though this may change in the near future currently no private
academic or research institutions in AA\&SP are operating in
Greece. Thus the individuals who are employed full time to teach or do
research in AA\&SP are typically civil servants in permanent, tenure
track, or fixed-term research associate positions. Currently the
majority of them (see Sect. 3, Fig. 1) are in Universities and only a
small fraction is associated with Research Institutes or
Observatories.

Up until the early 1980s the structure of the University system in
Greece followed the German style. It was based on ``Chairs'' of
Professors in specific research fields (i.e. Astronomy, Classical
Mechanics, etc.). The few astronomy Professors, typically as numerous
as the corresponding number of university departments pursuing
research in the various areas of AA\&SP, held their position until the
age of retirement. They made all major administrative decisions
related to both teaching priorities and research directions in their
institutions. Several junior staff members did support them in these
activities but those members had only marginal control and rather
limited independence to pursue their own research directions.

The University system presently in place was put forward in 1983,
(with a few minor modifications over the past $\sim$25 years) and has a
structure similar to the current academic system of the United
States. There are two ranks of tenure track positions: Lecturer and
Assistant Professor, and two of tenured positions: Associate Professor
and Full Professor. A minimum of three years is required on each rank
before applying for promotion the next. Tenure is obtained upon
successful evaluation after spending three years at the level of
Assistant Professor. A university faculty can, in principle pursue
his/her own research direction, teach courses, and supervise graduate
students.

However, when this change in the academic system took place in the
early 1980's, there was no provision on the age distribution of the
faculty to be hired. As a result a large number of individuals who
were already affiliated with the universities at the time in junior
level -- the so-called ``Assistant'' appointments, automatically
obtained tenure at the rank of Lecturer upon the completion of their
PhD. Others, who already had a PhD, were considered for tenure at
higher ranks. The evaluation for this process though was often not
very strict and with criteria based mostly on social reasons or giving
a disproportional emphasis to the teaching responsibilities of the
faculty, rather than mostly based on their research background and
potential or relevance of the field to the future direction of modern
astrophysics. In addition, since most of the individuals who obtained
these positions were past graduate students of the same universities,
there was a disproportional hiring from ``within''. This phenomenon,
known as ``academic inbreeding'', was more prevalent in the older
institutions in Athens and Thessaloniki, which had the largest number
of staff at the time. Even today there are institutions in Greece
where well over 70\% of their permanent staff members are past alumni
who did their dissertation in the same institute and did not spend
more than a couple of years away from their alma matter before
obtaining permanent positions. It is beyond the scope of the present
document to discuss this phenomenon and the serious negative
consequences it has on both the quality of research performed and on
the opening of new research horizons in academic institutions. We
should note though that this phenomenon is not unique to Greece, as it
also appears for example in the French academic system and in
Korea\footnote{See: Science 18 December 1998: Vol. 282. no. 5397} ,
but it is practically absent in the United States. Greek universities
in the periphery of the country did not suffer much from this problem
for two reasons. Either they had not produced their own PhDs due to
their youth as institutions, or their faculty made a conscious
decision to have a broader perspective in their hiring process. For
example, in the Department of Physics of the University of Crete,
where the author is currently employed, only 1 out of the 33 faculty
members obtained his PhD from this institution. All these political
decisions had implications that continue to affect the evolution of
Greek astronomy, and academic system in general, well into the 21st
century.

Research Institutes in Greece have a similar structure to the
Universities, with also four ranks, which are loosely indicated as
Researcher-D, -C, -B and -A. Each researcher also has to remain in a
given rank for a minimum of three years and tenure is obtained upon
promotion from Researcher-C to Researcher-B.  Research Institutes can
not award academic degrees and as a result close collaboration with a
University is needed in order for a researcher to be able to
co-supervise students.

Public funding for development of infrastructures, direct support of
research in astronomy, or fellowships towards graduate studies in the
field, has been traditionally fairly limited. Such a low level support
is not restricted to Greek astronomy but it is also the case in most
disciplines. In 2004 Greece spent only 0.58\% of its Gross Domestic
Product (GDP) in R\&D, which brings Greece as a nation in the last
place among the 15 EU countries in this category, a position it holds
for the past 5 years. At over the same period the European Union (EU)
average was 1.95\%, more than 3 times higher\footnote{Source EUROSTAT
in http://europa.eu.int/comm/eurostat/}. As a result the
possibilities for Greek astronomers to join large international
collaborative projects, or just to obtain support to attend scientific
meetings outside Greece have been scarce. Even though the situation
has recently improved over the past decade, and the possibilities --
mostly via the financial and organizational support of the European
Union -- are more numerous, the effects of this low level national
funding can be seen in most indices quantifying the overall astronomy
scientific output from Greece. It is worth noting that Greece, which
joined the European Union as the 10th member in 1981, is still not a
member state of the European Southern Observatory (ESO) and only
joined the European Space Agency (ESA) in 2005.

\section{Governing bodies}

The policies that directly affect issues related to AA\&SP in Greece
are determined by the Ministry of Development, in particular the
General Secretariat for Research and Technology, and the Ministry of
Education. The first has administrative control over the research
institutes and astronomy infrastructure and the latter controls the
national university system.

Greek astronomers can express their opinion or shape policies on
issues related AA\&SP via the Greek National Committee for Astronomy
(GNCA) or the Hellenic Astronomical Society (Hel.A.S.).

\subsection{The Greek National Committee for Astronomy (GNCA)}

The Greek National Committee for Astronomy (GNCA\footnote{The web page
of GNCA is: http://www.astro.noa.gr/$\sim$gnca}) was established, by Royal
Decree, as the official advisory committee to the Greek Government for
all matters relevant to Astronomical and Astrophysical research, in
1957. It is the official body, responsible for the promotion and
coordination of Astronomy in Greece and for all matters related to
international astronomical cooperation. The Minister of Development
selects the members of GNCA and appoints them for a term of two
years. Its official seat is the National Observatory of Athens. Since
1995, GNCA does not have its own budget, but obtains its funding from
the budget of the General Secretariat of Research and Technology
(GSRT) of the Ministry of Development. The GNCA has the following
principal objectives:

\begin{itemize}

\item To co-ordinate and promote the various astronomical activities
in Greece, including research and education.

\item To act as the link between the Greek astronomical community and
the International Astronomical Union (IAU), officially representing
Greece in the General Assembly of the IAU.

\item To facilitate the advancement of international collaboration
between Greek and foreign astronomers and research groups.

\end{itemize}

Besides the IAU, GNCA has taken responsibility for Greece's
representation to the Board of Directors of the journal "Astronomy and
Astrophysics" (and its financial contributions), to the European Joint
Organization for Solar Observations (JOSO) and, recently, to the
European Union FP6, I3, Network OPTICON (see Sect. 8).

The board of GNCA consists of five ordinary and five substitute
members. The current (2005-2007) ordinary members are
Prof. P. Laskarides (Chair - Univ. of Athens), Prof. T. Krimigis (Vice
Chair - Academy of Athens), Dr. I. Daglis (Nat. Obs. of Athens),
Prof. N. Kylafis (Univ. of Crete), and Prof. J.H. Seiradakis (Univ. of
Thessaloniki). The substitute members over the same period are:
Prof. S. Avgoloupis (Univ. of Thessaloniki), Dr. E. Dara (Academy of
Athens), Prof. M. Kafatos (George Mason Univ., USA), Prof. A. Nindos
(Univ. of Ioannina), and Prof. J. Papamastorakis (Univ. of Crete).

\begin{figure*} 
\plotone{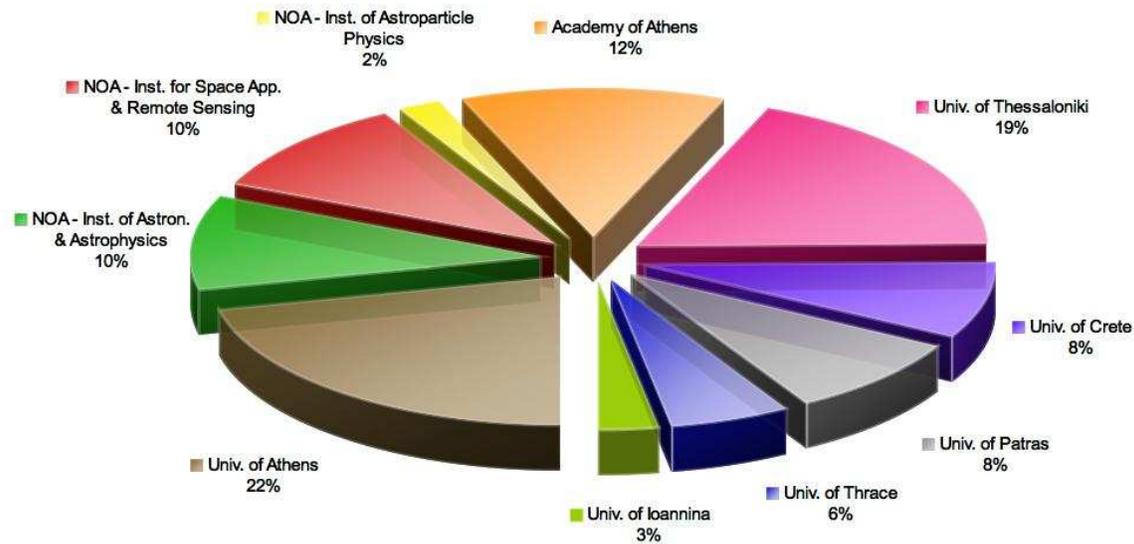}
\figurenum{1} 

\caption{Distribution of tenured and tenure track astronomers in the
major research institutes in Greece. The total number of astronomers
included in this study is 107.}

\end{figure*} 

\subsection{The Hellenic Astronomical Society (Hel.A.S.)}

The Hellenic Astronomical Society (Hel.A.S.\footnote{The web page of
Hel.A.S. is http://www.astro.auth.gr/elaset}) exists for nearly 15
years and it is the major association of professional astronomers in
Greece. Its overall structure and operation is similar to other
national societies such as the ``American Astronomical Society'' in
the US, or the ``Soci\'{e}t\'{e} Fran\c{c}aise d' Astronomie et
d'Astrophysique'' in France.

Historically, the first serious attempt to establish a Hellenic
Astronomical Society was undertaken in 1982 during the XVIII General
Assembly of the International Astronomical Union, which took place in
Patras, Greece. There, during several meetings, a dozen astronomers
gathered in order to put the foundations of the long sought
Society. The following years progress was slow even though material
necessary for setting up the framework for the Society was being
collected.  It was much later, in November 1991, when
Prof. P. Laskarides (Univ. of Athens) issued the first announcement of
the 1st Hellenic Astronomical Conference, that the idea of the
establishment of an Astronomical Society was formally put forward
again. With the help of several colleagues Prof. J.H. Seiradakis
(Univ. of Thessaloniki) drafted the first Constitution for the
Society. The final version was presented to the participants of the
1st Hellenic Astronomical Conference, which was held in Athens in
September 1992.

During the Athens Conference, several astronomers became founding
members of the Hellenic Astronomical Society. A few more founding
members signed the Constitution during the next weeks bringing the
total number of founding members to sixty six (66). Following the
appropriate legal procedures, the Hellenic Astronomical Society
(Hel.A.S.) was recognized by the Court of Justice in Athens on May 25
1993. The appointed Council of Hel.A.S. became aware of the verdict of
the Court of Justice in June 1993. The President of the Council,
Prof. B. Barbanis (Univ. of Thessaloniki), assisted by the members
initiated the procedure for the first elections of Hel.A.S. In the
elections, which took place on June 2nd 1994, participated 83\% of the
founding members.

According to its Constitution the Governing Council of
Hel.A.S. consists of a President, six (6) members and three (3)
auditors. The Council is elected for a two-year term and an individual
cannot serve on it for more than two consecutive terms. The candidates
for the Council must be members of Hel.A.S. who live and work
permanently in Greece during the term of their candidacy and at least
42\% of them should be affiliated with institutions outside the Athens
metro area. The current Council, whose mandate ends in June 2006,
consists of Prof. P. Laskarides (Univ. of Athens) as the president and
Prof. D. Hatzidimitriou (Univ. of Crete), Prof. K. Tsinganos (Univ. of
Athens), Prof. V. Geroyannis (Univ. of Patras), Prof. K. Kokkotas
(Univ. of Thessaloniki) and Prof. X. Moussas (Univ. of Athens) as
members. The auditors for the same period are Prof. E. Danezis
(Univ. of Athens), Prof. E. Mavromichalaki (Univ. of Athens) and
Dr. E. Xilouris (National Obs. of Athens)

The Hellenic Astronomical Society has been very active and currently
has 272 members, 27\% of which live and work outside Greece. It has
been recognized as an Affiliated Member of the European Astronomical
Society (EAS) and has established links with other international
astronomical societies. It has been organizing a major science meeting
every two years and in the summer of 1997 organized the Joint European
and National Astronomical Meeting (JENAM-97).

\begin{figure*} 
\plotone{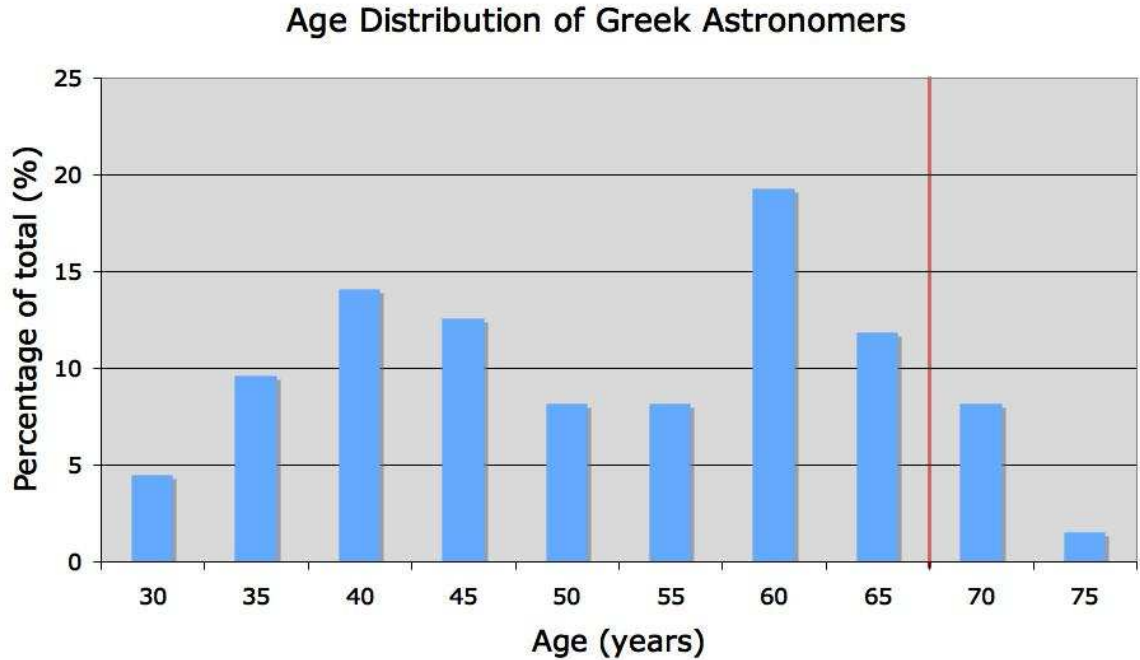}
\figurenum{2} 
\caption{A histogram of the age distribution of astronomers in Greece
in 2006. The vertical line indicates the 67th year of age which is the
current compulsory retirement age for civil servants.}

\end{figure*} 

\section{Academic Institutions}

\subsection{Human Resources in Astronomy, Astrophysics \& Space Physics}

As one would expect, since more than half of the population in Greece
is concentrated in the Athens and Thessaloniki metro areas, most of
the astronomers in Greece are also associated with institutes located
in these two cities. This is depicted in Figure 1 where the fraction
of tenured and tenure track astronomy faculty in the major AA\&SP
institutions in Greece is presented. 

An additional issue, which affects the current state and has direct
implications to the future of Greek astronomy, is related to the age
distribution of Greek professional astronomers. In Figure 2, a
histogram of 135 astronomers working in Greece is presented, using the
database of the members of the Hellenic Astronomical Society, as well
as ancillary information collected by the author. The study was
limited to individuals over the age of 30, since this is typically the
age when one is competitive for tenure track or long term research
associate positions. Some individuals over the retirement age of 67,
who are on an emeritus-type position and/or still active, were
included in the analysis. The error on a single 5-year bin is of the
order of 5\% but it is very likely that the values of bins at ages
greater than 55 are somewhat underestimated. This is due to the fact
that there are a number of individuals who formally have a tenured
astronomy positions but as they are no longer active they were not
included in the database of Hel.A.S, on which analysis was based.

Inspection of Figure 2 clearly reveals that almost 30\% of Greek
astronomers are near or over the age of 60. This was a direct
consequence of the legislative changes that took place in Greece in
the early 1980s mentioned in Sect. 1. Furthermore, statistics over the
last 10 years indicate that on average there were less than 3 new
tenure track astronomy position openings per year in the country,
including both universities and research institutions. The fraction of
astronomers near the age of retirement is even larger if we were to
consider only the two older universities of Athens and
Thessaloniki. This implies that within the next 5 to 10 years a large
number of their current faculty members will retire and they will have
to be replaced in a very short time scale. This will be an interesting
challenge for Greek astronomy. Will it be possible for these
institutions to find enough, well qualified, candidates from the
available pool of post-docs and research associates for their needs?
Will they be forced to lower their hiring standards in order to hire
faculty for their teaching needs, or they will be able to hire with a
lower pace, being selective and identifying the key scientific
research areas they should be investing in? In 2016 we will know the
answer to these questions!

Another topic worth touching upon is gender diversity in Greek
astronomy. At the time of writing this report 13\% of the permanent or
tenure track astronomy positions in Greece were held by women. This
percentage is less than in France\footnote{The percentages for the
other countries mentioned are based on the 2003 report by Dr. Florence
Durret (Institute d'Astrophysique de Paris, France) available at:
http://www2.iap.fr/sf2a/courrier.html}, which leads the way with
$\sim$26\%, or in Italy, Russia and Spain, all above 15\%, but higher
than the fraction of female astronomers in the United States which is
$\sim$10\%. We should note though, that only recently one female
astronomer in Greece reached for the first time the highest possible
academic rank (Full Professor or Researcher A), a statistic that will
hopefully improve very soon.

\subsection{Research in Astronomy, Astrophysics \& Space Physics}

The latest organized effort to map the research activity in AA\&SP in
the various institutes in Greece took place in 1998. Dr. E. Kontizas,
as the president of GNCA at the time, appointed an international
six-member committee, chaired by Prof. Y. Terzian (Cornell Univ.,
USA), to report on the status of astronomy in Greece and propose
recommendations for the future. The report\footnote{The complete
``Terzian Report'' is available at:
http://www.astro.noa.gr/gnca/NEWS/ca-report2000.htm} was presented
during the workshop ``Astronomy 2000+: Greek Prospects for the 21st
Century'' which took place at the National Observatory of Athens on
November 1998. The description presented in the following paragraphs
draws from material included in this report with some modifications
mostly related to changes in the human resources of the institutes
involved.

There are eight institutions in Greece with Departments or Sections
devoted to teaching and research in Astronomy and Astrophysics. Three
are located in Athens: the largest is the Section within the
Department of Physics of the National Kapodistrian University,
followed by the National Observatory of Athens, and an astronomy
Section of the national Academy of Athens. In Thessaloniki there is a
very small group within the Faculty of Engineering (Polytechnic
School) and a considerably larger one within the Department of Physics
of the Aristotle University. In Crete there is a Section of
Astrophysics and Space Physics in the Department of Physics in
Heraklion, while the Universities of Patras and Ioannina each have
small Astronomy groups within either their Physics or Engineering
Departments. Some research activity in very specific areas
(i.e. cosmology or general relativity) also exists in a few
Departments of Mathematics but the numbers of permanent staff are very
small and there is no critical mass to be considered groups.

The principal institutions devoted to research and technical
development in space sciences is the Department of Electrical and
Computer Engineering at the ``Democritus" University of Thrace (in
particular the Laboratory of Space Electrodynamics in the Section of
Telecommunications and Space Science) and the Institute for Space
Applications and Remote Sensing of the National Observatory of
Athens. Significant research in ground based ionospheric and
atmospheric work is also a component of the overall Astrophysics and
Space Science Section at the University of Crete. Activity relating to
space science also exists in the Section of Astronomy, Astrophysics,
and Mechanics of the University of Athens, and at the Research Center
for Astronomy in the Academy of Athens.

In the following subsections we present a brief description of the
various institutes in Greece hosting research groups with active
research in AA\&SP. More detailed annual activity reports from most
institutions and groups are being collected by the Greek National
Committee for Astronomy and they are made available from its web site
mentioned in section 2.1.

\subsubsection{University of Athens}

The ``National \& Kapodistrian'' University of Athens was founded in
1837, soon after the independence of Greece. It was the first
University in Greece as well as in the Balkan Peninsula and the whole
eastern Mediterranean region. The Department of Physics was created in
1904 and its current Section of Astronomy, Astrophysics, and Mechanics
was formed in the mid 1980s by merging the previously independent
Chairs indicated in its name. The Section is the largest in Greece and
consists of 24 tenured or tenure tract faculty. Most of them have
research interests in the area of Astronomy and Astrophysics, while
Mechanics is a rather small constituent. The Department of Physics
started a graduate school in 1994, within which the Section has its
own Masters and PhD programs with 12 graduate courses. Since January
2000 the Section also operates a 40cm Cassegrain telescope within a 5m
rotating dome located on the top of the Physics building. The
telescope was constructed by DFM engineering (USA) has an f/3 focal
ratio and it is mainly used for educational activities.

In addition to the pursuit of astronomy, it should also be mentioned
that the faculty of the Physics Department have been involved over
many years in building a deep sea High Energy Neutrino telescope,
known as NESTOR. Recently this effort has been put under the auspices
of the National Observatory of Athens as an independent institute for
Astroparticle Physics (see Sect. 3.2.2).

\subsubsection{National Observatory of Athens}

The National Observatory of Athens (NOA\footnote{More information on
the National Observatory of Athens the can be found at:
http://www.noa.gr}) was founded in 1842 and is the oldest research
institute in Greece. It currently consists of five institutes three of
which, the Institute of Astronomy and Astrophysics, the Institute for
Space Applications and Remote Sensing and the Institute of
Astroparticle Physics -- Nestor, conduct research in AA\&SP. The
current director of NOA is Prof. C. Zerefos.

The Institute of Astronomy and Astrophysics has 11 permanent staff
scientists as well as research associates and support personnel. Their
research interests include a variety of topics in extragalactic
astronomy, observational cosmology, interstellar matter, X-ray
astronomy, and binary stars. The Institute supports the Astronomical
Observatory in Kryoneri as well as the new Chelmos Observatory where
the new 2.3m ``Aristarchos'' telescope, the largest in Greece, is
located (see Sect. 4.1, 4.3). The institute is also very active in
public outreach activities, among which are the operation of a Visitor
Center and an annual summer school, which introduces basic concepts of
modern astrophysics to high-school students since 1996. The current
director of the Institute is Prof. C. Goudis.

The Institute for Space Applications and Remote Sensing has 11 tenure
or tenure track research staff. The activities of the Institute
encompass a wide area in Space Research and Applications. Its main
objective is to carry out R\&D projects in these fields, which include
Remote Sensing, Telecommunications, Space and Ionospheric
Physics. Additional activities include the systematic collection and
processing of data derived from observations made either from the
earth or space as well as the performance of autonomous studies in
other specific subjects of space research and applications. The
Institute is equipped with satellite and ionospheric ground stations,
various RF and electronic test and measurement equipment, as well as
an advanced computing center connected to international networks. The
current director of the Institute is Dr. I. Daglis.

The Institute of Astroparticle Physics -- NESTOR (Neutrino Extended
Submarine Telescope with Oceanographic Research) became the fifth
institute of the national Observatory of Athens in 2003. The institute
is leading the development of a deep-sea high energy neutrino
telescope approximately 14km off the shore from the town of Pylos in
Peloponnese, at water depth of 4000m. NESTOR will detect the Cherenkov
radiation produced by muons traversing the water when their parent
neutrinos emitted from astrophysical objects, such as X-ray binaries,
black holes, or Active Galactic Nuclei, interact with water. The
current director of the Institute is Prof. L. Resvanis.

\subsubsection{Academy of Athens}

The Academy of Athens was formally founded in 1926. It currently has
among its members two Academicians (Prof. G. Contopoulos and
Prof. T. Krimigis) with a background and research interests in
astronomy. One of the centers of the Academy, the Research Center for
Astronomy and Applied Mathematics, consists of 11 permanent research
staff, and conducts research in solar and space physics, cosmology,
particle physics and dynamical astronomy.

\subsubsection{University of Thessaloniki}

The ``Aristotle'' University of Thessaloniki was the second university
in Greece and it was founded in 1925. There are two units in the
University with activity in Astronomy. The smallest, in the
Polytechnic School, consists of two faculty members and their research
is concentrated mainly on flare stars. The largest is the Section of
Astrophysics, Astronomy and Mechanics (AAM\footnote{The online
description of the AAM Section in Thessaloniki can be found at:
http://www.astro.auth.gr}) of the Department of Physics, with 17
faculty members, several research associates, graduate students, and
support personnel. The Section was formed in the mid-80s when the
administrative structure of the Laboratories of Astronomy (founded in
1943) and Mechanics changed (see Sect. 1). The staff is active in many
areas of theoretical and observational astrophysics, including an
active theoretical group on gravitation and general relativity, as
well as in education and public outreach. In addition to the
Stephanion Observatory (see Sect. 4.4) the Section operates a 20cm
refracting telescope (made by Secretan, Paris) in a rotating
6m-diameter dome, which is located within the University campus, and
it is used for educational purposes.

\subsubsection{University of Crete}

The University of Crete was founded in 1973 but accepted its first
students in 1978. Its Department of Physics was founded in 1978 and is
the youngest of similar Departments in Greece. The Section of
Astrophysics and Space Physics\footnote{The web page of the Astronomy
Section in Crete can be found at: http://www.physics.uoc.gr/en/} has 7
faculty members as well as several research staff and graduate
students. Two (2) more tenured track astronomers, from the Foundation
for Research and Technology-Hellas and the Technical Education
Institute of Heraklion, are actively collaborating with the members of
the Section. Research at the Univ. of Crete covers a broad range in
theoretical and observational problems related to both galactic and
extragalactic astrophysics. Significant efforts are being devoted to
the operations of an Ionospheric Physics laboratory. Observations for
several astronomical projects are also taken at the Skinakas
Observatory (see Sect. 4.2) and others are performed using
international ground and space born telescopes. The Department has a
graduate program through which students can pursue their graduate
studies in astrophysics.

\subsubsection{University of Thrace}

The Laboratory of Space Electrodynamics (LSE) at Department of
Electrical and Computer Engineering of the ``Democritus" University of
Thrace is the largest space physics group in Greece with extensive
experience in hardware development. It consists of 6 faculty, several
research associates, support personnel, and many graduate and
undergraduate students. The scientists are co investigators or
associated scientists on several international spacecraft missions
(e.g. Ulysses, Geotail, Cluster II, and others), successfully funded
through European programs and bilateral collaborations with other
countries, including the U.S. The LSE has designed, developed and
successfully flown particle experiments on a number of Russian
spacecraft, as well as component systems to instruments involving data
processing units and ASICs (Application Specific Integrated
Circuits). Such high technology hardware capability in space
instrumentation is rather unique within Greece. The LSE group has
expanded their activities to antennae and propagation, satellite
communications, and other related areas.

\subsubsection{University of Patras}

The University of Patras has a Laboratory of Astronomy and a Section
of Astronomy in the Division of Theoretical and Mathematical Physics
in the Department of Physics. A total of nine tenured and tenure track
faculty teach courses and conduct research in a few astronomy areas
and there is an active theoretical group on celestial mechanics.

\subsubsection{University of Ioannina}

This is the smallest group of Astronomy in a Department of Physics in
Greece. It has three faculty members in the Section of Astrogeophysics,
within the Department of Physics. The staff performs research mostly
in solar physics and in multi-wavelength observations of flare stars.

\begin{figure*} 
\plotone{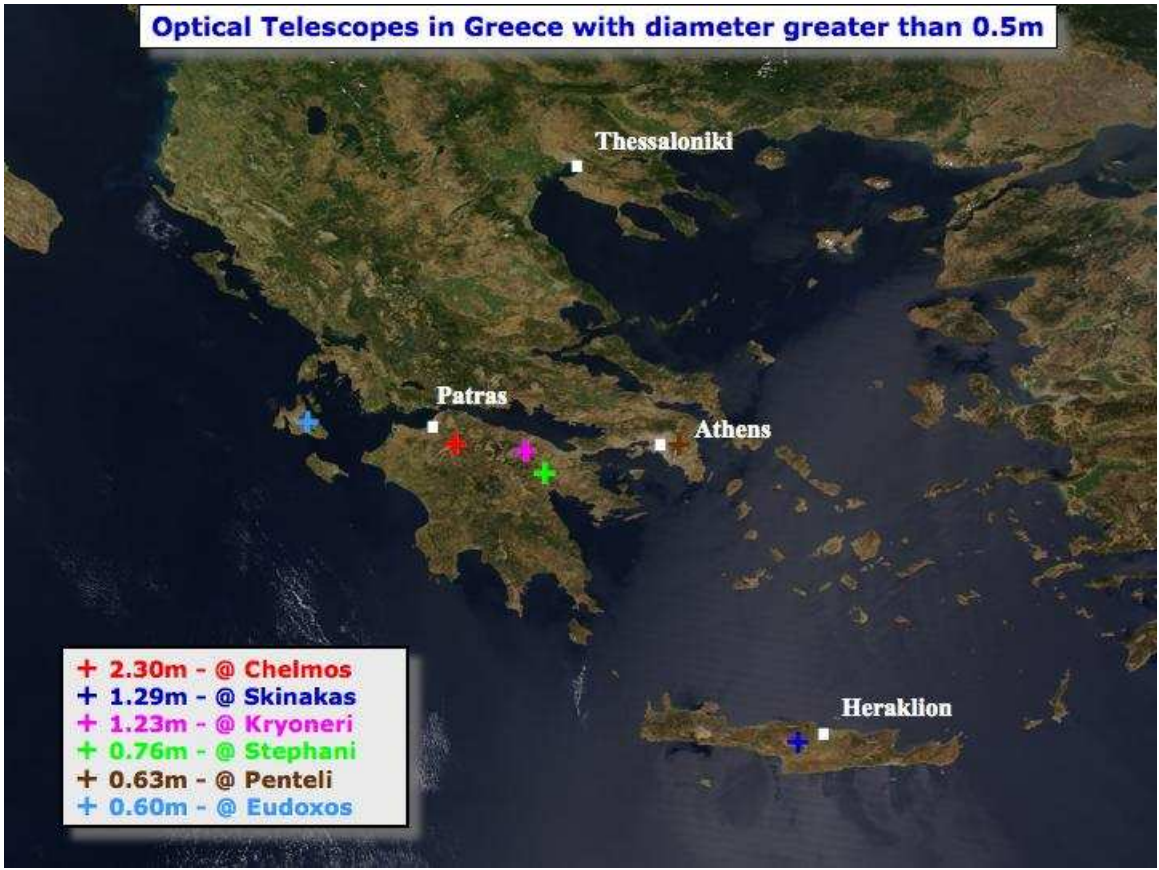}
\figurenum{3} 

\caption{An optical satellite image of Greece, obtained in 2004, in
which the locations of the major observatories hosting functioning
optical telescopes with a primary mirror diameter larger than 0.5m are
indicated (Image courtesy of MODIS Rapid Response Project at
NASA/GSFC)}

\end{figure*} 
 
\section{National Facilities}

The limited funding of the Greek government towards basic and applied
research has had, as a result, the small investment in major
infrastructures for astronomical facilities in Greece. This affected
the oldest observatories in Greece, such as Penteli, Kryoneri and
Stephanion Observatory, which have difficulties keeping up-to-date
with the modern developments in telescope design, aperture size of the
telescope primary mirrors, as well as the instrumentation
available. More recent facilities, such as Skinakas Observatory, which
currently hosts the largest operational telescope in Greece which is
1.29m in diameter, are more modern and do provide high quality
instruments to the observers. However, they also suffer from the
limited national financial support and they cannot function as
facilities that can provide access to all Greek astronomers who may
wish to use them. A major effort in improving the current situation
has been the ongoing construction of the 2.3m ``Aristarchos" telescope
by the National Observatory of Athens. The telescope had its first
light in the end of 2005 and when it becomes fully operational, before
the end of 2007, will be the largest in Greece.

\subsection{Chelmos Observatory}

The site selected for the new 2.3m telescope is located in Northern
Peloponnese, on top of Chelmos mountain, near the small town of
Kalavrita approximately 150km from Athens, with longitude:
22$^{o}$13', latitude: 37$^{o}$58' N and an elevation of 2340m. The
total cost for the project is expected to be about 5 million Euros and
it was financed mainly by the European Union, as well as by the
General Secretariat for Research and Technology of the Ministry of
Development. The telescope named ``Aristarchos" is a
Ritchey-Chrtien with a focal ratio f/8 and a 10Õ field of view as
well an RC-corrected field of view of 1degree. The telescope and dome
are constructed by Carl Zeiss (Germany).
 
\begin{figure*}[h] 
\plotone{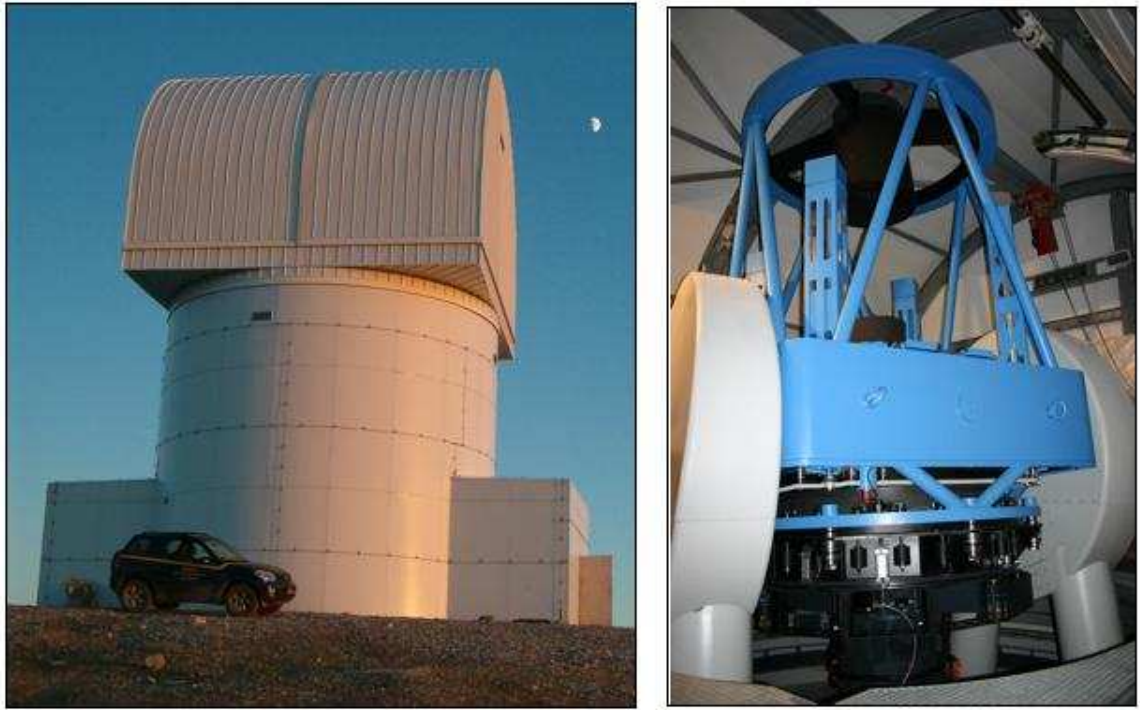}
\figurenum{4} 
\caption{Left: A picture of the dome of the 2.3m Aristarchos
telescope at Chelmos. Right: A photograph of the telescope inside the
dome (Images courtesy of National Observatory of Athens).}
\end{figure*}

The image scale on the focal plane is 1''=85$\mu$m and a 1024x1024 CCD
camera was the first light instrument. A medium resolution (2.5\AA -
6\AA) spectrometer covering the range between 4270\AA and 7730\AA as
well as a 4096x4096 optical CCD will be the first generation
instruments of the telescope. These will be followed by an echelle
spectrometer covering the range between 3900\AA and 7500\AA with a
resolution of 6 km s$^{-1}$, as well as other instruments. The
supervision of the telescope construction as well as its operation are
managed by the Institute of Astronomy and Astrophysics of the National
Observatory of Athens\footnote{Details on the Chelmos Observatory is
available at: http://www.astro.noa.gr/ASC\_2.3m/ngt\_main.htm}.

\subsection{Skinakas Observatory}

The Skinakas Observatory\footnote{More information on Skinakas
Observatory can be obtained from: http://skinakas.physics.uoc.gr}
operates as part of a scientific research collaboration between the
University of Crete, the Foundation for Research and Technology-Hellas
(FORTH) and the Max-Planck-Institut f\"{u}r Extraterrestrische (MPE)
Physik of Germany.

The site of the Observatory (Longitude: 24$^{o}$53'57''E, Latitude:
35$^{o}$12'43''N), chosen on scientific and functional grounds, is the
Skinakas summit of Mount Ida (also known as Psiloritis), at an
altitude of 1750m and a distance of 60km from Heraklion. The
Observatory has two telescopes: a Modified Ritchey-Chrtien
telescope with a 1.29m aperture (focal ratio f/7.6), which became
operational in 1995, and a 30cm telescope (focal ratio f/3.2). The
building for the small telescope was constructed in 1986, and
observations started in 1987. The site is one of the best in Greece
with weather conditions often permitting photometric sub-arcsecond
seeing. It includes a modern guest house powered with solar arrays and
an Internet connection.

\begin{figure*} 
\plotone{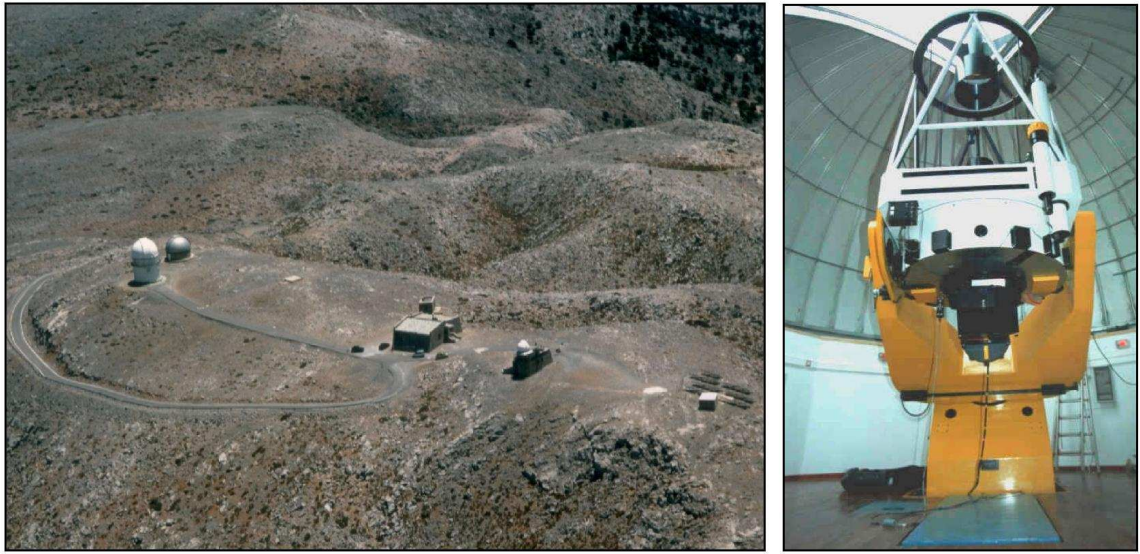}
\figurenum{5} 

\caption{An areal view of the Skinakas Observatory summit with the
larger dome of the 1.29m telescope seen on the left, along with the
smaller domes the guest house facilities. The 1.29m telescope inside
its dome is seen in the right (Images courtesy of the Physics Dept.,
Univ. of Crete).}

\end{figure*}

The optical system of the 1.29m telescope were manufactured by Carl
Zeiss (Germany). The mechanical parts were built by DFM Engineering
(USA). The instrumentation available includes a focal reducer, a
number of optical CCD cameras, and a low resolution long slit
spectrograph. A 1024x1024 near-IR camera and an echelle spectrograph
will soon be available on site, along with OPTIMA, a fast
photo-polarimeter with microsecond time resolution intended for
observations of compact objects. Various research projects, both
galactic and extragalactic, mostly led by members of the Department of
Physics of the University of Crete or MPE astronomers have been
ongoing since the facilities became operational. The close
collaboration with the MPE group and the FORTH engineering support has
helped the astrophysics group in Crete in keeping the telescope and
the instruments in the forefront of technology, always taking into
account the limitations in the budget.

\subsection{Kryoneri Observatory}

The Astronomical Station of Kryoneri\footnote{Additional information
on Kryoneri Observatory are at:
http://www.astro.noa.gr/ASK\_1.2m/ask\_main.htm} was established in
1972. It is located in the Northern Peloponnese, on top of mountain
Kilini at an elevation of 930m, near the small village Kryoneri 110km
from Athens (longitude: 22$^{o}$37'E, latitude: 37$^{o}$58'N). The
1.2m Cassegrain Coude telescope of the Astronomical Station Kryoneri,
made by Grubb Parsons Co., Newcastle, was installed in 1975. Its
optical system consists of a paraboloidal primary mirror of 1.23m in
diameter and f/3 focal ratio, and a hyperboloidal secondary mirror (31
cm). Both mirrors are made of Zerodur. The telescope focal ratio is
f/13, its field of view is about 40' and the image scale is
12.5"/mm. As with Chelmos Observatory, Kryoneri is operated by the
Institute of Astronomy and Astrophysics of the National Observatory of
Athens.

\subsection{Stephanion Observatory}

The first observations at the Stephanion Observatory, in eastern
Peloponnese, were undertaken in March 1967 with a guest 38cm reflector
and a UBV photometer that belonged to the Bergedorf Observatory of the
University of Hamburg, Germany. Since then a large number of
instruments have been hosted at the 800-m altitude observatory, which
is located at longitude: 22$^{o}$49'45''E, latitude: 37$^{o}$45'9''N,
including French telescopes, for monitoring satellites, and a 40cm
reflector from the Utrecht Observatory, Netherlands. In June 1971, the
30-inch (76cm) Cassegrain reflector of the University of Thessaloniki
was installed at the Observatory. Until 1975, when the 1.23m
Cassegrain Coude reflector at Kryoneri became operational, this was
the largest telescope in Greece.

The 30-inch reflector is mounted asymmetrically and its focal ratio is
f/3 for the primary hyperbolic mirror and f/13.5 for the Cassegrain
focus. It was constructed by Astro Mechanics, USA, a firm that has
long ago discontinued making astronomical instruments. The majority of
observations are carried out with a Johnson dual channel photoelectric
photometer with an offset guider unit mounted in the Cassegrain
focus. It includes an RCA 1P21 and an RCA 7102 photo-multipliers, both
of which are refrigerated by dry ice. Key photometric observations of
variable stars (flare stars, Cepheid variables, RS CVns, etc) have
been undertaken in co-operation with large ground or space
instruments. The international demand for co-operative and
simultaneous observations at the Stephanion Observatory stems from the
strict differential method used for obtaining absolute, above
atmosphere, stellar magnitudes in the international UBV system. The
error in the calibrated magnitudes obtained is usually better than
0.02 magnitudes.

\subsection{Penteli Observatory}

The Astronomical Station on Penteli Mountain, just 15km from downtown
Athens, was established in 1937 when it became apparent that it was
necessary to move the telescopes from the grounds of the old National
Observatory in the center of Athens. In 1955 the National Observatory
of Athens accepted the donation offered by the University of
Cambridge, for a 62.5cm telescope designed by R. S. Newall, and
constructed by the firm Thomas Cooke \& Sons in 1868. Its big tube
(about 9m in legth), the German-type equatorial mount and its weight
of about nine tons, required careful dismounting, transportation and
installation in a new dome that was built in Penteli. This telescope,
no longer used for research, is still available on site today.

\subsection{Eudoxos Educational Observatory}

The ``Eudoxos" observatory is a web-accessible complex of optical and
radio telescopes, founded in 1999, whose facilities are located 16km
from Argostoli, in the Ionian island of Kefallinia at a plateau 600m
below the peak of mount Ainos (1628m). It operates a 0.6m Cassegrain
robotic telescope named after Dr. Andreas Michalitsianos, a Greek
astrophysicist who was born in the island and had a successful career
in NASA (USA) until his early passing away. The observatory was formed
by a consortium of Greek institutes involving the National Research
Center of Physical Sciences ``Democritus", the Hellenic Naval Academy,
the Ministry of Education and the Prefecture of Kefallinia and
Ithaki. It is being operated by the same consortium with the addition
of the University of Athens and has already received substantial
support from the Hellenic Air Force and the Ministry of Education. The
0.6m telescope consists of a fully autonomous computerized optical
tube assembly, automated enclosure, GPS smart antenna for time
synchronization, a full set of meteorological sensors, a large format
imaging CCD camera and UBVRI wheeled photometric filters, as well as a
fleet of peripheral instruments currently under construction or
testing. All equipment is completely controlled by two supervisory
computers, which communicate via the Internet to the participating
secondary schools and institutions.

\section{High School and Undergraduate Studies in Astronomy}

The Greek secondary education system does provide substantial training
in physics and mathematics to the students who wish to follow
university studies in sciences. Even though there is no compulsory
astronomy course in high school (only an elective introductory
astronomy course is available for high-school juniors) basic astronomy
ideas related to the solar system, stars, galaxies, and the formation
of the universe are presented in other courses. Since 1996 the
``Society for Space and Astronomy" of Volos (see Sect. 7) has been
organizing a very successful national astronomy competition in which
students from all over Greece attending the last three years of high
school (``Lyceum'' in greek) can participate.  The top students are
awarded various prizes while the first two are invited to attend an
all-expenses-paid summer space-camp in the United States organized by
NASA. This effort, mainly supported by private funds and volunteer
work, has helped substantially in popularizing astronomy among high
school students.

At the university level there is no Bachelors (BSc) degree in
Astronomy or Space Science in Greece. Most individuals, who are now
professionals in the field of AA\&SP and did their undergraduate
studies in Greece, obtained their BSc degree in Physics following a
four-year program. Some, mostly theorists, have obtained their
undergraduate degrees in Mathematics or Engineering. Even in the
various Departments of Physics in Greece though, the curriculum of the
astronomy courses varies depending on the number and research
background of the astronomy faculty. Most Physics majors in Greece
have to follow at least one compulsory junior course in Astrophysics
while some complementary topics on dynamical astronomy are typically
covered on compulsory sophomore and junior level courses in classical
mechanics and modern physics. Most Departments of Physics offer the
possibility of an astronomy specialization (or minor), even though
this is not formally awarded as a degree. Within this framework,
students, who are interested in astronomy, have the opportunity to
attend typically five to ten junior and senior level courses in
astrophysics, space physics and celestial mechanics, thus obtaining a
fairly solid background if they wish to continue for graduate studies.

The level of this University astronomy training is usually very good
in the theoretical and encyclopedic part and the top students are
competitive with international standards. What the students lack
sometimes is the hands-on practical knowledge, which can only be
obtained with access to engineering facilities or observatories. The
organization of summer schools, such as the one taking place at the
University of Crete for the past 17 years, addressed to undergraduates
at junior and senior level, can often fill this gap. There are also
recent efforts at various institutions, such as the Univ. of Athens,
to enhance the observational astrophysics courses with a more
organized usage of small telescopes and new instruments.

\section{Graduate Studies in Astronomy, Astrophysics \& Space Physics}

Graduate studies in Astronomy Astrophysics \& Space Physics leading to
a Masters or a PhD degree can now be completed in most Greek
Universities. The first well-organized physics graduate program in
Greece with coursework, qualifying exams, and at least partial
financial support for students was developed in the University of
Crete in 1984. This was soon to be followed by the University of
Athens and other institutions.

However, the system suffers from difficulties, which again stem from
the limited national funding. Less than a handful of state fellowships
for graduate studies in AA\&SP are available each year. Providing
financial support for graduate studies via European Union or national
research proposals in astronomy is very challenging both due to
limited funds available in this field as well as due to various
bureaucratic and organizational difficulties. As a result graduate
students in Greece have to either work, or rely on other means to
support themselves during their studies. This sometimes affects their
ability to invest the amount of time necessary for research in order
to complete a very high quality PhD project.

These reasons have been pushing many of the Greek students to go
abroad for their graduate studies. The most popular destinations are
the United States, the United Kingdom, Germany, France and The
Netherlands. The improved facilities and competitive research
environment in those countries do provide high quality training to the
students but often decrease the likelihood of their return to work in
Greece. Recently though, the opportunities made available by the
European Union, mostly via the Human Capital as well as Training and
Mobility programs, have ameliorated the situation. These new
possibilities have provided the means to establish close links between
Greek and other European institutions, which improves substantially
the training of local students thus bringing direct scientific return
to the home institution.

\section{Amateur Astronomy in Greece}

Amateur astronomy has been flourishing in Greece over the past
decade. The availability of high quality and low cost small telescopes
and the use of Internet to organize and advertise the activities of
groups has greatly helped the development in the field. Many amateur
organizations exist all over Greece. In particular one should mention
the ``Hellenic Astronomical Union" which is the society of amateur
astronomers in Athens, the ``Group of friends of Astronomy" in
Thessaloniki, the ``Corfu Astronomical Society", and the very active
``Society for Space and Astronomy" in the city of Volos.  Since 1999
the Greek amateur astronomers have been organizing a national meeting
every two years where they present their results and discuss issues of
common interest.

\section{Greece and International Astronomy Organizations}

Greece joined the International Astronomical Union as a funding member
in 1920.  It also contributes to the support of the international
refereed journal of Astronomy \& Astrophysics, which allows Greek
astronomers to publish their scientific results without page charges.
Since 2004 Greece also participates in OPTICON, a 19.2 million Euro
5-year European FP6 Infrastructure Network, which provides access to a
number of medium size telescope facilities around the world.

In early 2005 Greece joined the European Space Agency (ESA)
contributing to the annual budget of ESA with $\sim$9 million
Euros. This opens new opportunities for Astrophysics and Space Physics
both in terms of technology development as well as in science. Over
the past year significant organizational efforts have been taking
place in order to stimulate the Greek AA\&SP community so that it will
be able to capitalize on this investment and join the rest of the
western European countries in the forefront of space technology.

\section{Remarks}

I believe that it is appropriate to end this article on the status of
Greek astronomy with an optimistic note on the many improvements we
have all experienced over the past decade. As it can be seen from the
material presented in the previous sections, the environment, both
research and academic, for the current and next generation of Greek
astronomers is considerably better than what our predecessors had
experienced and worked through.

I must also touch upon, a subject, which was mentioned earlier but
only briefly. Unfortunately Greece is still not a member of the
European Southern Observatory, the major astronomical organization in
Europe. As a result it has no access to the current European
infrastructures of the Very Large Telescope (VLT), nor to the
development of the Atacama Large Millimeter Array (ALMA) nor to the
design of the future ESO projects, such as the 100m OverWhelmingly
Large telescope (OWL). I should stress that the report\footnote{The
complete ``Terzian Report'' is available at
http://www.astro.noa.gr/gnca/NEWS/ca-report2000.htm} of the
international expert committee chaired by Prof. Terzian (Cornell
Univ.) on the status of Greek Astronomy presented in 1998 during the
workshop ``Astronomy 2000+: Greek Prospects for the 21st Century''
noted that joining ESO should be the first astronomy priority for the
nation. Current rough estimates indicate that the cost for Greece to
join ESO would be a one-time $\sim$10 million Euros entrance fee,
similar to our annual contribution to ESA, and an annual membership
fee of only $\sim$1 million Euros. Thus, if following the
recommendation of the international expert committee, ESO were to be
our lofty astronomy goal for the present century, one can only hope
that the whole Greek community will embrace it and with a joined
effort will convince the ``powers that be" to turn the wheels and make
it a reality before the end of the current decade.

\acknowledgments

This document used material from the online archives of the Hellenic
Astronomical Society, the Greek National Committee for Astronomy, as
well as from the annual reports of the various institutes that were
available online. I would like to thank K. Kokkotas (Univ. of
Thessaloniki), N. Kylafis (Univ. of Crete), J.H. Seiradakis (Univ. of
Thessaloniki), K. Tsinganos (Univ. of Athens), and E. Xilouris
(National Observatory of Athens) for making suggestions that improved
this article.

\end{document}